\documentclass{PoS}

\title{Bound H-dibaryon from Full QCD Simulations\\ on the Lattice}

\ShortTitle{Bound H-dibaryon from Full QCD Simulations on the Lattice}

\author{\speaker{Takashi Inoue}\\
        Nihon University, College of Bioresource Sciences, 
        Fujisawa 252-0880, Japan\\
        E-mail: \email{inoue.takashi@nihon-u.ac.jp}}

\author{
for HAL QCD Collaboration
}
\author{
\begin{center}
\includegraphics[width=0.33\textwidth]{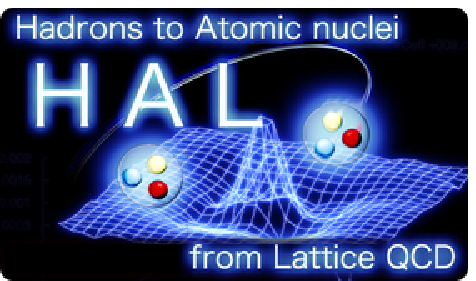}
\end{center}
}

\abstract{
 Using a new method recently proposed by the HAL QCD collaboration to study hadron interactions in lattice QCD,
 we investigate a possibility for an existence of the $H$-dibaryon in the flavor $SU(3)$ symmetric world.
 A potential for the flavor-singlet baryon-baryon channel is derived from the Nambu-Bethe-Salpeter wave function, 
 and the elusive $H$-dibaryon is shown to be a bound state, 
 with the binding energy of 20--50 MeV for the pseudo-scalar meson masses in the range that 469--1171 MeV.
}

\FullConference{ The XXIX International Symposium on Lattice Field Theory - Lattice 2011\\
July 10-16, 2011\\
Squaw Valley, Lake Tahoe, California}

\begin{document}

\section{Introduction}

Lattice QCD now became a powerful method to study not only hadron spectra and QCD phase structure, 
but also hadronic interactions and multi-hadron systems.
For example, hadron scatterings such as $\pi\pi$ and $NN$ have been studied by using the L\"uscher's finite volume method \cite{Luscher:1990ux,Fukugita:1994ve,Beane:2002nu}.
Recently, the binding energy of helium nucleus was also measured directly in a quenched lattice QCD \cite{Yamazaki:2009ua}.
 In 2007, an approach to study hadron interactions 
 on the basis of the Nambu-Bethe-Salpeter amplitude on the lattice was proposed\cite{Ishii:2006ec,Aoki:2009ji}: This approach (the HAL QCD method) has been
 successfully applied to  baryon-baryon interactions (such as $NN$ and $YN$)
and the meson-baryon interaction (such as $\bar K N$)\cite{Nemura:2008sp,Inoue:2010hs,Ikeda:2010sg}.  

 In any of the above approaches, 
the spatial lattice size $L$ should be large enough
 to accommodate multi-hadrons inside the lattice volume. 
Once  $L$ becomes large, however, energy levels of hadrons corresponding to the scattering state in the lattice box become dense,
so that isolation of the ground state from the excited states becomes very difficult
 unless unrealistically large imaginary-time $t$ is employed.
Recently, we proposed a new technique to resolve this issue by generalizing the original HAL QCD method \cite{Ishii:2011,Inoue:2010es}.
With our new method, information of hadron interactions can be obtained even without separating energy eigenstates on the lattice.

An interesting application of the above new method is the elusive $H$-dibaryon
which has been know to receive large finite volume effect on the lattice \cite{Wetzorke:2002mx}.
The $H$-dibaryon, predicted by R.~L.~Jaffe in 34 years ago~\cite{Jaffe:1976yi},
is one of most famous candidates of exotic-hadron. 
The prediction was based on the observations that the Pauli exclusion principle 
can be completely avoided due to the  flavor-singlet($uuddss$)
nature of $H$-dibaryon, together with the large attraction from
one-gluon-exchange interaction between quarks~\cite{Jaffe:1976yi,Sakai:1999qm}.
Search for the $H$-dibaryon is one of the most challenging theoretical and experimental problems
in the physics of strong interaction. 
Although deeply bound $H$-dibaryon with the binding energy $B_H > 7 $ MeV from the $\Lambda\Lambda$ threshold
has been ruled out by the discovery of the double $\Lambda$ hypernuclei,
$_{\Lambda \Lambda}^{\ \ 6}$He~\cite{Takahashi:2001nm}, there still  remains a possibility of 
a shallow bound state or a resonance in this channel~\cite{Yoon:2007aq}.
 
In this report, we first explain our new method to resolve the difficulty of the finite volume effect.
By using the method, we then study the $H$-dibaryon in lattice QCD.
To avoid unnecessary complications, we consider the flavor $SU(3)$ limit in this study.
We find a bound $H$-dibaryon with the binding energy of 20--50 MeV for the pseudo-scalar meson mass of 469--1171 MeV. 
 Comparison of our result with 
   those by other groups \cite{Luo:2007zzb,Beane:2010hg} as well as  
possible  $H$-dibaryon with flavor $SU(3)$ breaking are also discussed.

\section{Formalism}

In the original works \cite{Ishii:2006ec,Aoki:2009ji}, 
the Nambu-Bethe-Salpeter (NBS) wave function for the two-baryon system with  energy $E$, 
\begin{equation}
 \phi_E(\vec r, t)
   = \sum_{\vec x} \langle 0 \vert B_i(\vec x + \vec r,t)B_j(\vec x,t) \vert B=2, E \rangle,
\label{eqn:NBS}
\end{equation}
has been employed to study baryon-baryon ($BB$) interactions.
A correlation function $\Psi(\vec r,t)$  for two baryons can be expressed in terms of  $\phi_E(\vec r, t)$ in a finite volume as
\begin{equation}
 \Psi(\vec r, t)
   \, = \,  A_{\rm gr} \phi_{E_{\rm gr}}(\vec r)e^{-E_{\rm gr}\,t} ~ + ~ A_{\rm 1st}\phi_{E_{\rm 1st}}(\vec r) e^{-E_{\rm 1st}\,t} ~ \cdots
\label{eqn:tdepNBS}
\end{equation}
where $E_{\rm gr}$ and $E_{\rm 1st}$ are energy of the ground state and the first excited state,
respectively, and $A_{\rm gr}$ and $A_{\rm 1st}$ are the corresponding coefficients. 
Hereafter we call $\Psi(\vec r, t)$ a time($t$)-dependent NBS wave function. 
In principle, $\Psi(\vec r, t)$ is saturated by the ground state contribution
for $(E_{\rm 1st}-E_{\rm gr})\times t \gg 1$,  so that the wave function of the ground state can be extracted.
If the ground state is a scattering state or a weakly bound state,
the energy differences $E_{i}-E_{j}$ are relatively large (several hundred MeV) in a small volume  (e.g. $L\simeq 2$ fm), 
and hence the condition can be fulfilled at relatively small $t$ ( $t\simeq 1.2$ fm)
which we can access in actual lattice simulations.

In a large volume, however, much larger $t$ is required.
\begin{figure}[t]
\centering
  \includegraphics[width=0.475\textwidth]{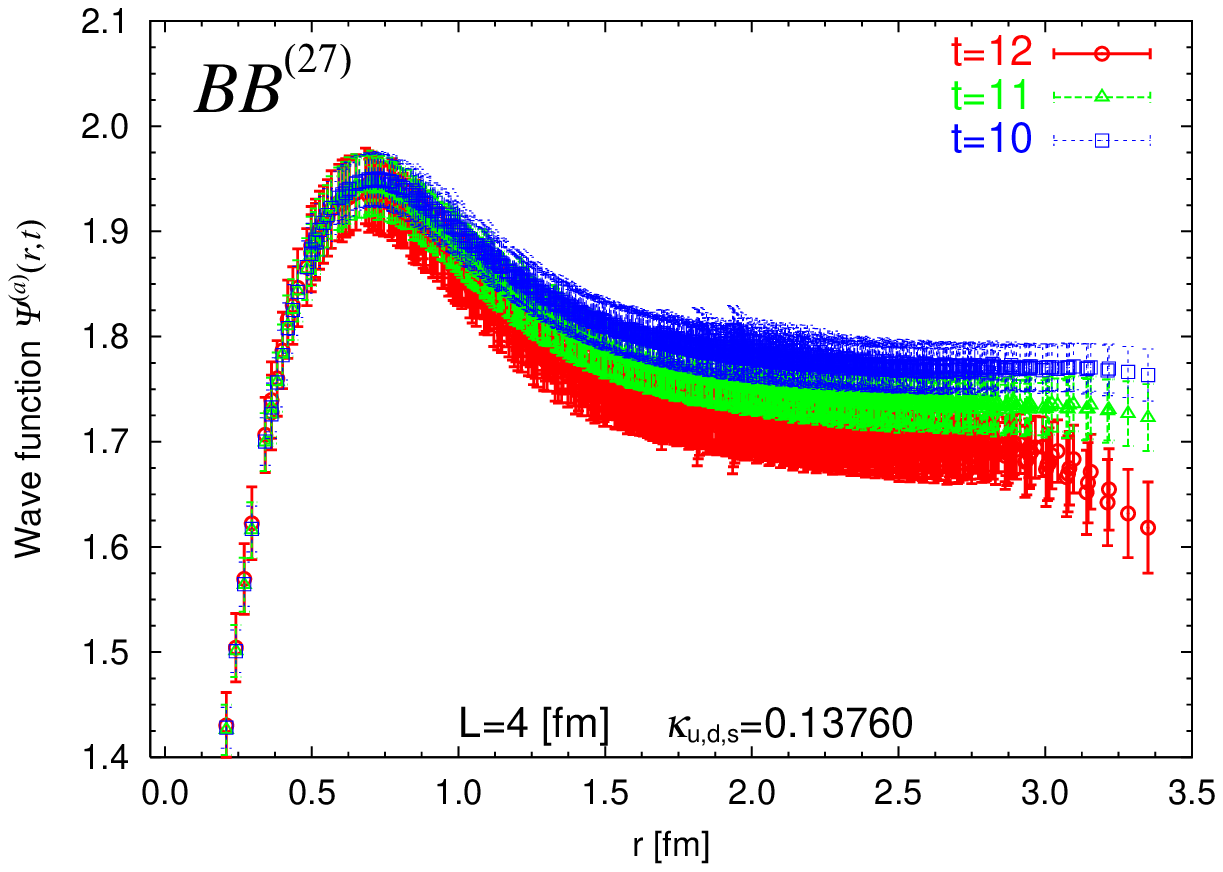}\hfill
  \includegraphics[width=0.475\textwidth]{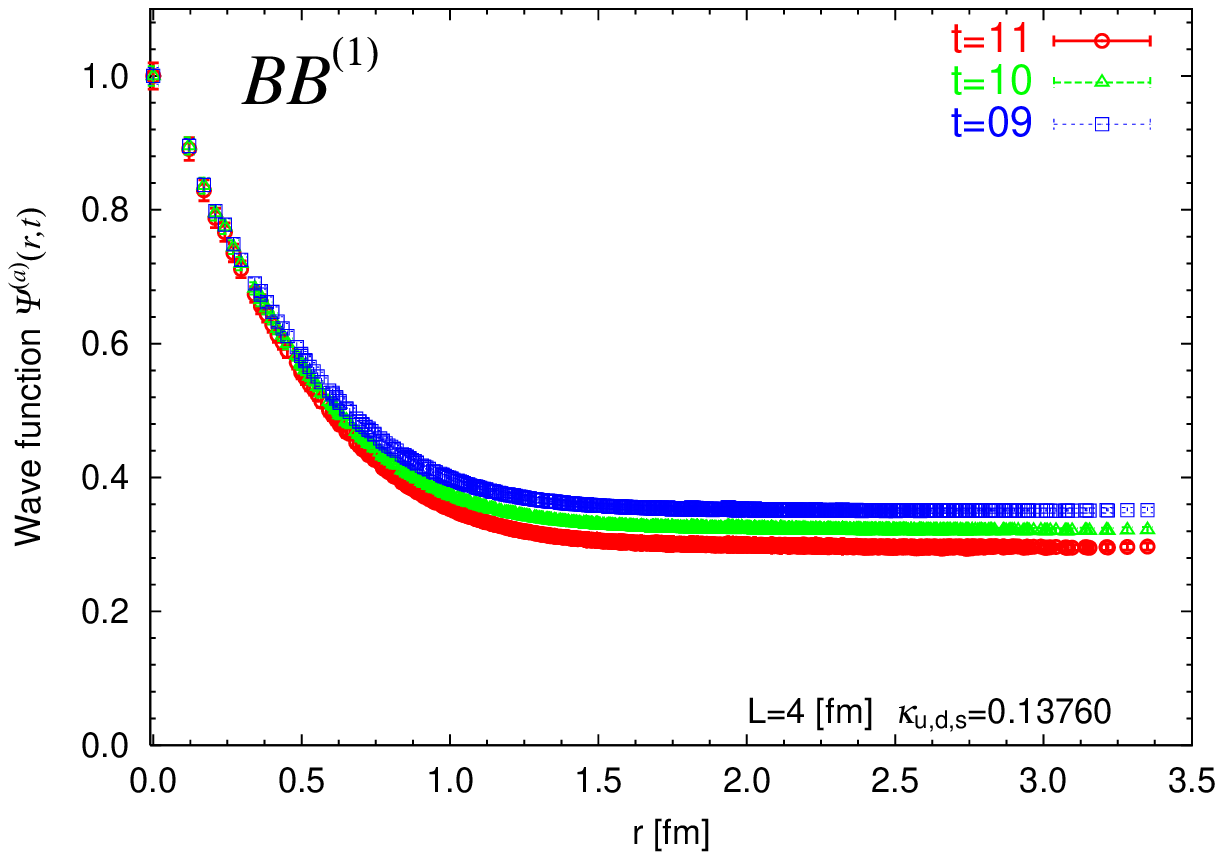}
\caption{Example of Nambu-Bethe-Salpeter wave function of two-baryon system measured on  the lattice with $L \simeq 4$ fm. 
Left (Right) panel shows the function for two-baryon in the flavor 27-plet (singlet) channel.
All data are normalized to unity at the origin.}
\label{fig:NBS}
\end{figure}
Fig. \ref{fig:NBS} shows the $t$-dependent NBS wave function of two-baryon
measured in our lattice QCD simulation on $L \simeq 4$ fm lattice, 
at sink-time $t$ around 10 in lattice unit ($a \simeq 0.12$ fm).
Shapes of the function change as the sink-time $t$ increases from 10 to 12, showing that 
$\Psi(\vec r, t)$ is not saturated by the ground state contribution in this range of $t$.
Since the energy difference tends to decrease as $L^{-2}$( with assumption that a deeply bound state is absent), $t$ around $40$ in lattice unit ($t\simeq 5$ fm in physical unit) may be necessary for the ground state saturation. 
We need huge statistics to extract signals at such large $t$, however,
since  the signal to noise ration of $\Psi(\vec r, t)$, which includes 4 baryon operators,
becomes bad as $t$ increases.  Such a calculation would be unacceptably expensive even with today's computational resources.
We therefore conclude that it is practically impossible to achieve the ground state saturation
for the two-baryon system in large volume, unless we employ some sophisticated techniques such as
the variational method with optimized sources.

To overcome this difficulty in large volume, the HAL QCD collaboration has recently proposed alternative method \cite{Ishii:2011,Inoue:2010es} as summarized below.
Within the non-relativistic approximation,
wave functions we consider satisfy the Schr\"{o}dinger equation with non-local but energy-independent potential $U(\vec r, \vec r')$.
For the lowest two energy eigenstates, it reads
\begin{eqnarray}
      \left[ M_1 + M_2 - \frac{\nabla^2}{2\mu} \right] \,\, \phi_{\rm gr}(\vec r)e^{-E_{\rm gr}\,\,t}
    \,\, + \, \int \!\! d^3 \vec r' \, U(\vec r, \vec r') \,\, \phi_{\rm gr}(\vec r)e^{-E_{\rm gr}\,\,t}
    &=& E_{\rm gr} \,\, \phi_{\rm gr}(\vec r) \, e^{-E_{\rm gr}\,\,t} \\
      \left[ M_1 + M_2 - \frac{\nabla^2}{2\mu} \right]  \phi_{\rm 1st}(\vec r) e^{-E_{\rm 1st}\,t}
    + \int \!\! d^3 \vec r' \, U(\vec r, \vec r') \, \phi_{\rm 1st}(\vec r) e^{-E_{\rm 1st}\,t}
    &=& E_{\rm 1st} \, \phi_{\rm 1st}(\vec r) e^{-E_{\rm 1st}\,t}
\end{eqnarray}
where $M_{1,2}$ represent masses of two baryons and $\mu$ is the reduced mass of the two baryons.
Since these equations (and those for other $E$) are linear in $\phi_E$, 
$\Psi(\vec r,t)=\sum_n A_n\phi_{E_n}({\vec r}) e^{-E_n t}$ satisfies 
\begin{equation}
      \left[ M_1 + M_2 - \frac{\nabla^2}{2\mu} \right] \Psi(\vec r, t)
    + \int \!\! d^3 \vec r' \, U(\vec r, \vec r') \, \Psi(\vec r', t) 
    = - \frac{\partial}{\partial t} \Psi(\vec r, t).
 \label{eq:t-dep}
\end{equation}
Using this equation, we can extract $U(\vec r, \vec r')$ from the $t$-dependent NBS wave function $\Psi(\vec r, t)$ at the moderate value of $t$,
which can be easily calculated in lattice QCD simulations. 
This is our new method to study hadron interactions in lattice QCD without isolating energy eigenstates.  
In practice, we expand the non-local potential $U(\vec r, \vec r')$ in terms of velocity such that
$U(\vec r, \vec r') = (V(\vec r)+O(\nabla))\delta^3(\vec r,-\vec r')$. 
In this paper, we only consider the leading term  $V(\vec r)$ of the velocity expansion. 

\begin{figure}[t]
\centering
  \includegraphics[width=0.475\textwidth]{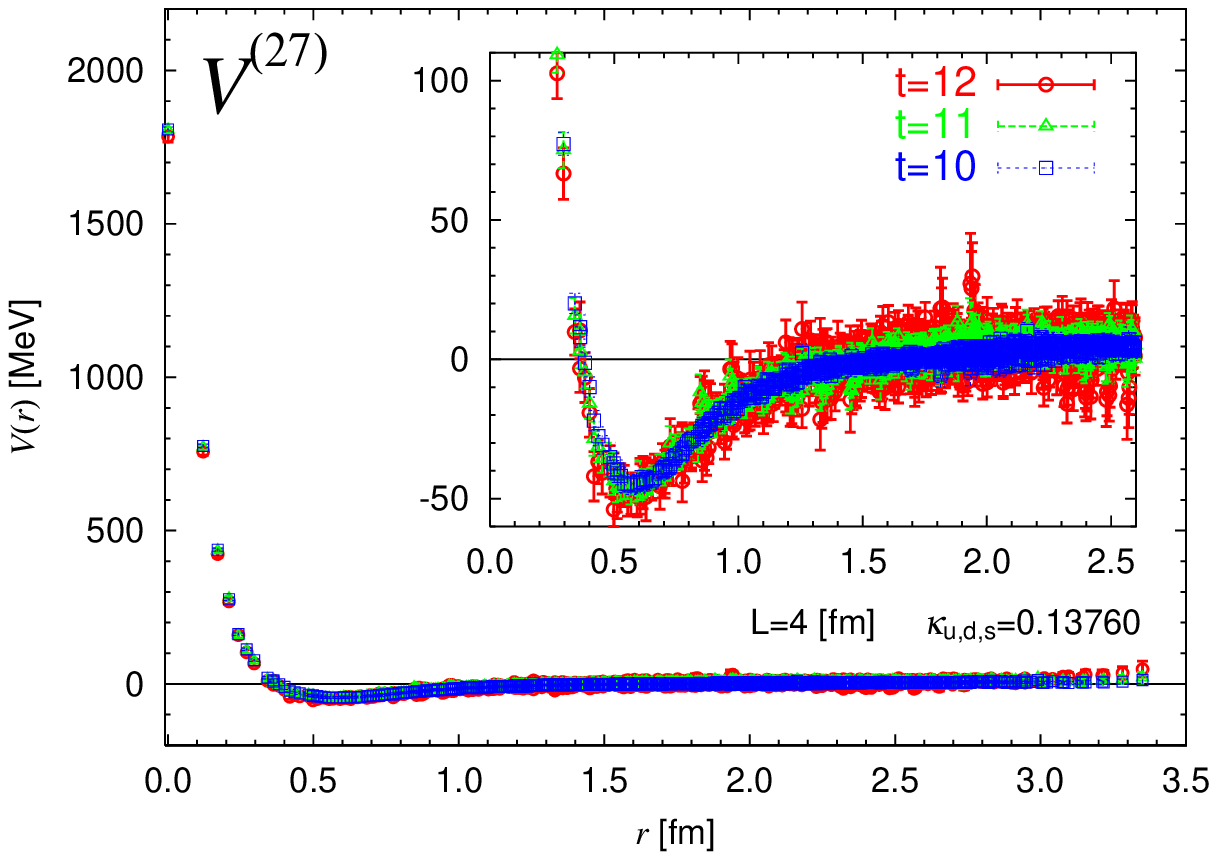}\hfill
  \includegraphics[width=0.475\textwidth]{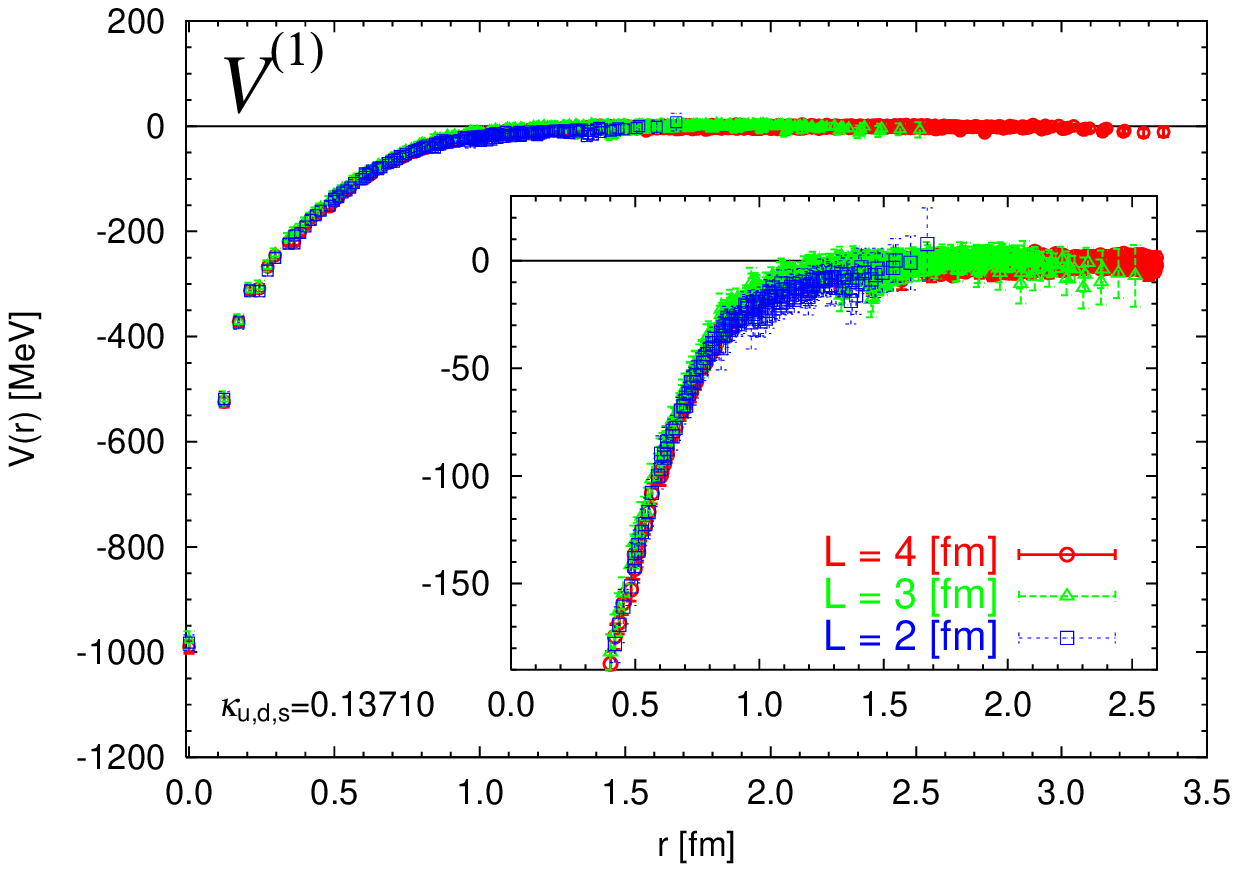}
\caption{Example of baryon-baryon potential extracted from QCD in our new method.
Left (Right) panel shows the potential for two-baryon in the flavor 27-plet (singlet) channel.
One can see independence of the potential on the sink-time $t$ and the volume $L^3$ in the left and the right panel, respectively.}
\label{fig:Indep}
\end{figure}

Fig. \ref{fig:Indep} shows examples of baryon-baryon potentials extracted by the new method.
In the left panel,  we show the spin-singlet and flavor 27-plet potential $V^{(27)}(r)$, which  is found to be independent of the sink-time $t$.
(Note that $V(\vec r)$ becomes  a function of  $r=\vert \vec r\vert$ for the spin-singlet sector.)
This result shows not only a success of the new method using eq.(\ref{eq:t-dep}) but also a goodness of the leading order approximation of the velocity expansion, since the $t$ independent potential $V(r)$ can  be extracted from the manifestly $t$ dependent function $\Psi(\vec r, t)$.
In the right panel,  the flavor singlet potential $V^{(1)}(r)$ is plotted on  $L=2,3$ and $4$ [fm] lattices.
Since the potential remains almost identical between $L=3$ and $4$ fm, 
the potential obtained on $L=4$ fm is considered to be volume independent.  
This observation is consistent with a fact that the interaction range of the potential, which is around $1.5$ fm from the figure, is smaller than the half of the lattice extension $L=4$ fm.
Once we obtain a volume independent potential,
any observable of the system, such as binding energy  and scattering phase shifts,
can be extracted by solving the Schr\"{o}dinger equation in the infinite volume.

\section{Lattice QCD setup}

\begin{table}[t]
\caption{Summary of lattice QCD parameters and hadron masses. See the main text for details.}
\label{tbl:mass}
\centering
 \begin{tabular}{c|c|c|c|c|c|c}
   \hline
  $a$ [fm] & $L$ [fm] &  $\kappa_{uds}$  & ~$M_{\rm p.s.}$ [MeV]~ &  ~$M_{\rm vec}$ [MeV]~& ~ $M_{\rm bar}$ [MeV]~ &
   ~$N_{\rm cfg}\,/\,N_{\rm traj}$~ \\
   \hline 
              &      &   ~0.13660~ &   1170.9(7) &   1510.4(0.9) & 2274(2) & 420\,/\,4200 \\
              &      &   ~0.13710~ &   1015.2(6) &   1360.6(1.1) & 2031(2) & 360\,/\,3600 \\
   {0.121(2)} & 3.87 &   ~0.13760~ & ~\,836.5(5) &   1188.9(0.9) & 1749(1) & 480\,/\,4800 \\
              &      &   ~0.13800~ & ~\,672.3(6) &   1027.6(1.0) & 1484(2) & 360\,/\,3600 \\
              &      &   ~0.13840~ & ~\,468.6(7) & ~\,829.2(1.5) & 1161(2) & 720\,/\,3600 \\

   \hline
 \end{tabular}
\end{table}

For dynamical lattice QCD simulations in the flavor $SU(3)$ limit,
we have generated ensembles of gauge configurations on a $32^3 \times 32$ lattice
with the renormalization group improved Iwasaki gauge action at $\beta=1.83$
and the non-perturbatively $O(a)$ improved Wilson quark action at five different value of quark mass. 
The lattice spacing $a$ is found to be 0.121(2) fm and hence lattice size $L$ is 3.87 fm.  
Hadron masses on each ensemble are given in  Table \ref{tbl:mass}, 
together with other parameters such as
the quark hopping parameter $\kappa_{uds}$,  
number of thermalized trajectory $N_{\rm traj}$ and number of configuration $N_{\rm cfg}$.
 
On each gauge configurations, 
the baryon two-point and four-point correlation functions are constructed from
quark propagators for the wall source
with the Dirichlet boundary condition in the temporal direction.
Baryon operators at source are combined to generate the two-baryon state in a definite flavor irreducible representation, while the local octet-baryon operators are used at sink.
To enhance signal, 16 measurements are made for each configuration,
together with the average between forward and backward propagation in time.
Statistical errors are estimated by the Jackknife method with bin size equal to 12 for the $\kappa_{uds}=0.13840$ and 6 for others.

\section{Results}

\begin{figure}[t]
\includegraphics[width=0.475\textwidth]{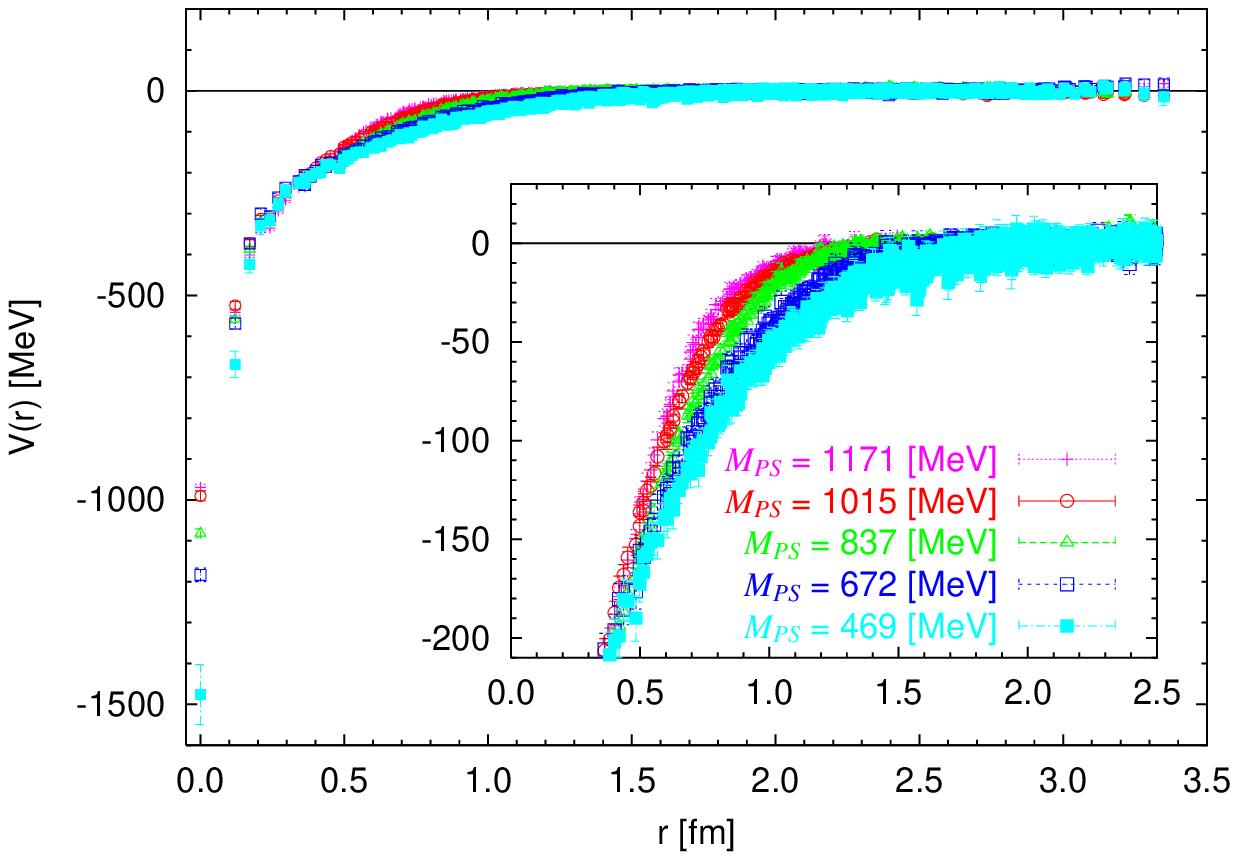} \quad
\includegraphics[width=0.475\textwidth]{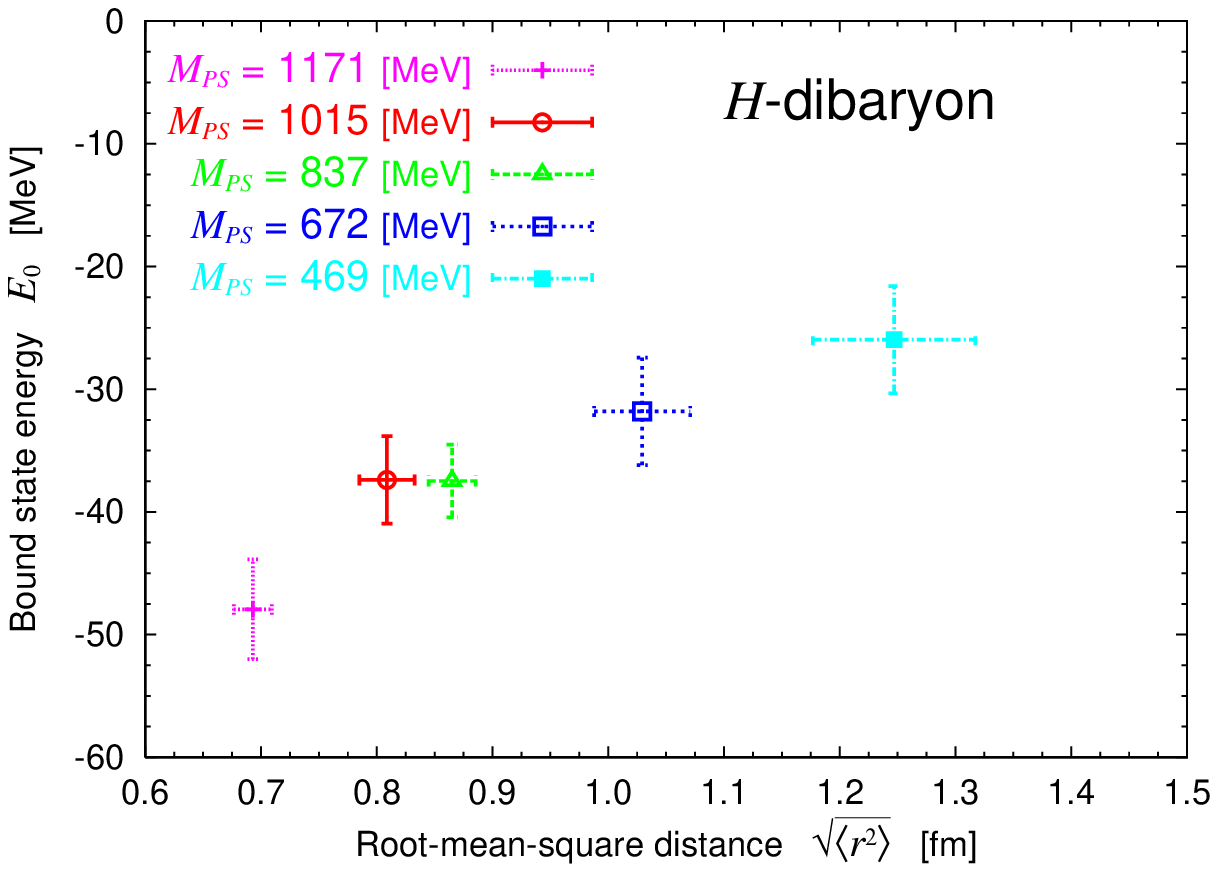}
\caption{ Left: Baryon-baryon potential in the flavor singlet channel. 
          Right: The ground state of the system. The energy $E_0$ is measured from two-baryon threshold.
          The bars indicate statistical error only.
}
\label{fig:V1andH}
\end{figure}

We here consider the system with baryon-number $B=2$ in the flavor-singlet and $J^P=0^+$ channel, to investigate $H$-dibaryon.
The left panel in Fig. \ref{fig:V1andH} shows the baryon-baryon potential in this channel at five values of quark mass: it  is entirely attractive and has an ``attractive core" \cite{Inoue:2010hs}.
This result is consistent with the prediction by Jaffe and also by the quark-model in
 this channel.
The figure also indicates that the attractive interaction becomes stronger as the quark mass decreases.

By solving the Schr\"{o}dinger equation with this potential, 
we have found one bound state in this channel~\cite{Inoue:2010es}.
The right panel in Fig. \ref{fig:V1andH} gives energy and size of this bound state, showing
that a stable $H$-dibaryon
exists at this range of the quark mass in the flavor $SU(3)$ symmetric world.

As shown in the right panel of Fig. \ref{fig:NBS},
the $t$-dependent NBS wave function for this channel goes to non-zero value at large distance,
contrary to a naive expectation for bound state wave functions.
The non-vanishing wave function at large distance is due to the contribution from excited states; they
 correspond to scattering states in the infinite volume and do not vanish at large distance.
In other words, the $t$-dependent  NBS wave function is a superposition of the bound state and scattering states.
Even in such a case, our new method is shown to work well.

The  binding energy ${\tilde B}_H = -E_0$ for the $H$-dibaryon ranges from 20 to 50 MeV and decreases as the quark mass decreases.
The rms distance $\sqrt{\langle r^2 \rangle}$ is a measure of a "size" of the $H$-dibaryon,
which may be compared to the rms distance of the deuteron in nature, $1.9 \times 2 = 3.8$ fm.
Although quark masses are different between the two,
this comparison suggests that $H$-dibaryon is much more compact than the deuteron.

By including a small systematic error caused by the choice of sink-time $t$ in the $t$-dependent NBS wave function,
the final result for the $H$-dibaryon binding energy become
\begin{eqnarray}
 M_{\rm p.s.}&=& 1171  ~\mbox{MeV} :~ \  {\tilde B}_H = 49.1 (3.4)(5.5) ~\mbox{MeV} \\
 M_{\rm p.s.}&=& 1015  ~\mbox{MeV} :~ \  {\tilde B}_H = 37.2 (3.7)(2.4) ~\mbox{MeV} \\
 M_{\rm p.s.}&=& ~~837 ~\mbox{MeV} :~ \  {\tilde B}_H = 37.8 (3.1)(4.2) ~\mbox{MeV} \\
 M_{\rm p.s.}&=& ~~672 ~\mbox{MeV} :~ \  {\tilde B}_H = 33.6 (4.8)(3.5) ~\mbox{MeV} \\
 M_{\rm p.s.}&=& ~~469 ~\mbox{MeV} :~ \  {\tilde B}_H = 26.0 (4.4)(4.8) ~\mbox{MeV} 
\end{eqnarray} 
with the statistical error (first) and the systematic error (second).
A bound $H$-dibaryon is also reported by the
 full QCD simulation with a different approach from ours \cite{Beane:2010hg}.
The obtained binding energy from the $\Lambda\Lambda$ threshold $B_H = 16.6(2.1)(4.6)$ MeV
at $(M_{\pi},M_{K})\simeq(389, 544)$ MeV is consistent with the present result.

A deeply bound $H$-dibaryon is ruled out by the discovery of the double $\Lambda$ hypernucleus.
The binding energy $\tilde{B}_H$ in this paper, however, should be interpreted as
the binding energy from the $BB$ threshold averaged in the $(S,I)=(-2,0)$ sector.
In the real world, the $\Lambda\Lambda$, $N\Xi$ and $\Sigma\Sigma$ thresholds in this sector largely split due to the flavor $SU(3)$ breaking.
We therefore expect that the binding energy of the $H$-dibaryon measured from the $\Lambda\Lambda$ threshold in nature
becomes much smaller than the present value or even the $H$-dibaryon goes above the $\Lambda\Lambda$ threshold.
Our trial calculation using a phenomenological flavor $SU(3)$ breaking indicates
that the $H$-dibaryon state becomes a resonance state above the $\Lambda\Lambda$ threshold but 
below the $N\Xi$ threshold, as the flavor $SU(3)$ breaking reaches the physical value.
To make a definite conclusion on this point, however, 
we need lattice QCD simulations at the physical point, together with the coupled channel analysis. 
We have already developed the formula for this purpose \cite{Aoki:2011gt} and tested the method in numerical simulations \cite{sasaki2010}.
A study towards this final goal is now in progress.

\medskip

\section*{Acknowledgments}
We thank K.-I. Ishikawa and PACS-CS group for providing their DDHMC/PHMC code~\cite{Aoki:2008sm},
and authors and maintainer of CPS++~\cite{CPS}, whose modified version is used in this paper.
Numerical computations of this work have been carried out at Univ. of Tsukuba supercomputer system (T2K).
This research is supported is supported by Grant-in-Aid for Scientific Research on Innovative Areas(No.2004:20105001, 20105003)
and for Scientific Research(C) 23540321.


\begin{thebibliography}{99}

\bibitem{Luscher:1990ux}
  M.~L\"{u}scher,
  Nucl.\ Phys.\  B {\bf 354}, 531 (1991).

\bibitem{Fukugita:1994ve}
  M.~Fukugita {\it et al.},
  Phys.\ Rev.\  D {\bf 52}, 3003 (1995)
  [arXiv:hep-lat/9501024].
\bibitem{Beane:2002nu}
  S.~R.~Beane and M.~J.~Savage,
  Phys.\ Lett.\  B {\bf 535}, 177 (2002)
  [arXiv:hep-lat/0202013].
\bibitem{Yamazaki:2009ua}
  T.~Yamazaki, Y.~Kuramashi, A.~Ukawa and f.~C.~Collaboration,
  Phys.\ Rev.\  D {\bf 81}, 111504 (2010)
  [arXiv:0912.1383 [hep-lat]].


\bibitem{Ishii:2006ec}
  N.~Ishii, S.~Aoki and T.~Hatsuda,
  Phys.\ Rev.\ Lett.\  {\bf 99}, 022001 (2007)
  [arXiv:nucl-th/0611096].
\bibitem{Aoki:2009ji}
  S.~Aoki, T.~Hatsuda and N.~Ishii,
  Prog.\ Theor.\ Phys.\  {\bf 123}, 89 (2010)
  [arXiv:0909.5585 [hep-lat]].


\bibitem{Nemura:2008sp}
  H.~Nemura, N.~Ishii, S.~Aoki and T.~Hatsuda,
  Phys.\ Lett.\  B {\bf 673}, 136 (2009)
  [arXiv:0806.1094 [nucl-th]].
\bibitem{Inoue:2010hs}
  T.~Inoue {\it et al.}  [HAL QCD Coll.],
  Prog. Theor. Phys. {\bf 124}, 591 (2010)
  [arXiv:1007.3559 [hep-lat]].
\bibitem{Ikeda:2010sg}
  Y.~Ikeda {\it et al.},
  arXiv:1002.2309 [hep-lat].


\bibitem{Ishii:2011}
  N.~Ishii {\it et al.}  [HAL QCD Coll.], in preparation
\bibitem{Inoue:2010es}
  T.~Inoue {\it et al.}  [HAL QCD Coll.],
  Phys.\ Rev.\ Lett.\  {\bf 106}, 162002 (2011)
  [arXiv:1012.5928 [hep-lat]].


\bibitem{Wetzorke:2002mx}
  I.~Wetzorke and F.~Karsch,
  Nucl.\ Phys.\ Proc.\ Suppl.\  {\bf 119}, 278 (2003)
  [arXiv:hep-lat/0208029].

\bibitem{Jaffe:1976yi}
  R.~L.~Jaffe,  
  Phys.\ Rev.\ Lett.\  {\bf 38}, 195 (1977)
  [Erratum-ibid.\  {\bf 38}, 617 (1977)].
\bibitem{Sakai:1999qm}
  T.~Sakai, K. Shimizu and K.~Yazaki,
  Prog.~Theor.~Phys.~Suppl. {\bf 137}, 121 (2000)
  [arXiv:nucl-th/9912063].
\bibitem{Takahashi:2001nm}
  H.~Takahashi {\it et al.},
  Phys.\ Rev.\ Lett.\  {\bf 87}, 212502 (2001).
 \bibitem{Yoon:2007aq}
  C.~J.~Yoon {\it et al.},
  Phys.\ Rev.\  C {\bf 75}, 022201 (2007).


\bibitem{Luo:2007zzb}
  Z.~H.~Luo, M.~Loan and X.~Q.~Luo,
  Mod.\ Phys.\ Lett.\  A {\bf 22}, 591 (2007)
  [arXiv:0803.3171 [hep-lat]].
\bibitem{Beane:2010hg}
  S.~R.~Beane {\it et al.}  [NPLQCD Coll.],
  Phys.\ Rev.\ Lett.\  {\bf 106}, 162001 (2011)
  [arXiv:1012.3812 [hep-lat]].

\bibitem{Aoki:2011gt}
  S.~Aoki {\it et al.}  [HAL QCD Collaboration],
  Proc. Jpn. Acad., Ser.B, {\bf 87}, 509 (2011)
  arXiv:1106.2281 [hep-lat].

 \bibitem{sasaki2010} 
  K.~Sasaki [HAL QCD Coll.], PoS {\bf LAT 2010}, 157 (2010).

\bibitem{Aoki:2008sm}
  S.~Aoki {\it et al.}  [PACS-CS Coll.],
  Phys.\ Rev.\  D {\bf 79}, 034503 (2009)
  [arXiv:0807.1661 [hep-lat]].

\bibitem{CPS}
Columbia Physics System (CPS), http://qcdoc.phys.columbia.edu/cps.html

\end{thebibliography}
\end{document}